\documentclass[aps,pre,twocolumn,showpacs,bibnotes,twoside,eqsecnum]{revtex4}
\usepackage{graphicx}
\usepackage{amsfonts}
\usepackage{amsmath}
\usepackage{amssymb}
\usepackage{multirow}
\usepackage{braket}
\usepackage[usenames,dvipsnames]{color}

\begin{document}
%\bibliographystyle{apsrev}

%\begin{frontmatter}

\title{Entropy and long-range correlations in DNA sequences}

\author{ S.~S.~Melnik\footnote[1]{melnikserg@yandex.ru} and O.~V.~Usatenko\footnote[2]
{usatenko@ire.kharkov.ua}}
\address{A. Ya. Usikov Institute for Radiophysics and Electronics \\
Ukrainian Academy of Science, 12 Proskura Street, 61805 Kharkov,
Ukraine}

\begin{abstract}

We analyze the structure of DNA molecules of different organisms by
using the additive Markov chain approach. Transforming nucleotide
sequences into binary strings, we perform statistical analysis of
the corresponding ``texts''. We develop the theory of $N$-step
additive binary stationary ergodic Markov chains and analyze their
differential entropy. Supposing that the correlations are weak we
express the conditional probability function of the chain by means
of the pair correlation function and represent the entropy as a
functional of the pair correlator. Since the model uses two point
correlators instead of probability of block occurring, it makes
possible to calculate the entropy of subsequences at much longer
distances than with the use of the standard methods. We utilize the
obtained analytical result for numerical evaluation of the entropy
of coarse-grained DNA texts. We believe that the entropy study can
be used for biological classification of living species.
\end{abstract}

\pacs{87.14.gk, 05.40.-a, 02.50.Ga} \maketitle

%\end{frontmatter}

\section{Introduction}

At present there is a commonly accepted viewpoint that our world is
complex and correlated. For this reason systems with long-range
interactions (and/or with long-range memory) and natural sequences
with non-trivial information content have been the focus of a large
number of studies in different fields of science over the past
several decades. Some of the most peculiar manifestations of this
concept are DNA and protein sequences~\cite{Buldyrev,Almir,mad}.

One of the efficient methods to investigate the correlated systems
is based on the decomposition of the space of states into a finite
number of parts labeled by definite symbols. This procedure,
referred to as a coarse graining, is accompanied by the loss of
short-range memory between states of system but does not affect and
does not damage its robust invariant statistical properties on large
scales. The most frequently used method of the decomposition is
based on the introduction of two parts of the phase space. In other
words, it consists in mapping the two parts of states onto two
symbols, say 0 and 1. Thus, the problem is reduced to investigating
the statistical properties of the symbolic binary sequences. This
method is applicable for the examination of both discrete and
continuous systems~\cite{Eren,Lind}.

There are many methods for describing the complex dynamical systems
and random sequences connected with them: correlation function,
fractal dimensions, multi-point probability distribution functions,
and many others. One of the most convenient characteristics serving
to the purpose of studying complex dynamics is the
entropy~\cite{Shan,Cover}. Being a measure of the information
content and redundancy in a sequence of data it is a powerful and
popular tool in examination of the complexity phenomena. It has been
used for the analysis of a number of different dynamical systems.

A standard method of understanding and describing statistical
properties of real physical systems or random sequences of data can
be represented as follows. First of all, we need to analyze the
sequence to find the correlation functions or the probabilities of
words occurring, with the length $L$ exceeding the correlation
length $R_{c}$ but being shorter than the length $M$ of the
sequence,
\begin{equation} \label{Strong}
 R_{c}<L<<M.
\end{equation}
At the same time, the number $d^L$ of different words of the length
$L$ composed in the alphabet containing $d$ letters has to be much
less than the number $M-L$ of words in the sequence,
\begin{equation}\label{Strong2}
 d^L<<M.
\end{equation}
The next step is to express the correlation properties of the
sequence in terms of the conditional probability function (CPF) of
the Markov chain, see below Eq.~(\ref{soglas}). Note, the Markov
chain should be of order $N$, which is supposed to be longer than
the correlation length,
\begin{equation}\label{Strong3}
R_{c}<N.
\end{equation}
This is the critical requirement because the correlation length of
natural sequence of interest (e.g., written or DNA texts) is usually
of the same order as the length of sequences. None of inequalities
(\ref{Strong})---(\ref{Strong3}) can be fulfilled. Really, the
lengths of words that could represent correctly the probability of
words occurring are 4-5 letters for a real natural text of the
length $10^6$ (written on an alphabet containing 27-30 letters and
symbols) or of order of 20 symbols for a coarse-grained text
represented by means of a binary sequence.

So, it is clear that the method described above can only describe
the random sequences with short correlation lengths and is not
suited for analyzing the systems with long-range correlations. The
latter issue will be the subject of our interest. We suppose that
all we need for constructing the sequence with long-range
correlations are the pair correlation functions.

We use the developed method~\cite{muya} for constructing the
conditional probability function presented by means of the pair
correlator which makes it possible to calculate analytically the
entropy of the sequence. It should be stressed that we suppose that
the correlations are weak but not short.

The scope of the paper is as follows. First, we discuss briefly
$N$-step additive Markov chain model~\cite{RewUAMM} and, supposing
that the correlations between symbols in the sequence are weak, we
express the conditional probability function by means of the pair
correlation function. In the next section we represent the
differential entropy in terms of the conditional probability
function of the Markov chain and express the entropy as the sum of
squares of the pair correlators. Then we discuss some properties of
the results obtained. Next, a fluctuation contribution to the
entropy due to finiteness is examined. The application of the
developed theory to some specific DNA sequences of nucleotides is
considered. In conclusion, some remarks on directions in which the
research can be progressed are presented.

\section{Additive Markov chains}

Consider a sequence $\mathbb{A}=a_{-\infty}^{\infty}= ...,
a_{-1},a_{0}, a_{1},...$ of real random variables $a_{i}$ taken from
the finite alphabet $A=\{1,2,...,d\}$, $a_{i}\in A$. The sequence
$\mathbb{A}$ is \textit{$N$-step Markov chain} (also referred to as
the higher-order or $N$-th-order Markov
chain~\cite{Raftery,Ching,Li,Cocho,Seifert}) if it possesses the
following property: the probability of symbol~$a_i$ to have a
certain value $a$ under the condition that the values of all
previous symbols are specified depends only on the values of $N$
previous symbols,
%
%\begin{eqnarray}\label{def_mark}
%P(a_i=a|\ldots,a_{i-2},a_{i-1})=P(a_i=a|a_{i-N},\ldots,a_{i-2},a_{i-1}).
%\end{eqnarray}
\begin{eqnarray}\label{def_mark}
&&P(a_i=a|\ldots,a_{i-2},a_{i-1})\\[6pt]
&&=P(a_i=a|a_{i-N},\ldots,a_{i-2},a_{i-1}). \nonumber
\end{eqnarray}
Sometimes the number $N$ is also referred  to as the \emph{memory
length} of the Markov chain.  The conditional probability function
(CPF) $P(a_i=a|a_{i-N},\ldots,a_{i-2},a_{i-1})$ determines
completely all statistical properties of the Markov chain and the
method of its iterative numerical construction. If the sequence,
whose statistical properties we would like to analyze is assigned,
the conditional probability function of the $N$-th order can be
found by a standard method,
\begin{equation}\label{soglas}
P(a_{N+1}=a|a_{1},\ldots,a_{N})=\frac{ P(a_{1},\ldots,a_{N},a) } {
P(a_{1},\ldots,a_{N})},
\end{equation}
where $P(a_{1},\ldots,a_{N},a)$ and $P(a_{1},\ldots,a_{N})$ are the
probabilities  of the $(N+1)$-word $a_{1},\ldots,a_{N},a$ and
$N$-word $a_{1},\ldots,a_{N}$ occurring, consequently.

The Markov chain determined by Eq.~(\ref{def_mark}) is a
\textit{homogeneous} sequence because its conditional probability
does not depend explicitly on $i$, i.e., is independent of the
position of symbols $a_{i-N},\ldots,a_{i-1},a_{i}$ in the chain. It
depends only on the values of $a_{i-N},\ldots,a_{i-1},a_{i}$ and
their positional relationship. The homogeneous sequences are
\emph{stationary}: the average value of any function
$f(a_{r_1},a_{r_1+r_2},\ldots , a_{r_1+\ldots+r_{s}})$  of $s$
arguments
\begin{eqnarray}\label{epsilon-av}
&& \overline{f}\,(a_{r_1},\ldots ,
a_{r_1+\ldots+r_{s}})\\[6pt]
&&=\lim_{M\to\infty}\frac{1}{M} \sum_{i=0}^{M-1} f(a_{i+r_1},\ldots
, a_{i+r_1+\ldots+r_{s}}). \nonumber
\end{eqnarray}
depends on $s-1$ differences between the indexes. In other words,
all statistically averaged functions of random variables are
\emph{shift-invariant}.

We assume that the chain is \emph{ergodic}. According to the Markov
theorem (see, e.g., Ref.~\cite{shir}), this property is valid for
the homogenous Markov chains if the strict inequalities,
\begin{equation}\label{ergo_m}
 0 < P(a_i=a|a_{i-N}^{i-1}) < 1, \quad i \in \mathbb{Z} = ...,-1,0,1,2,...
\end{equation}
are fulfilled for all possible values of the arguments in function
(\ref{def_mark}). Hereafter we use the shorter notation
$a_{i-N}^{i-1}$ for $N$-word $a_{i-N},...,a_{i-1}$. It follows from
the ergodicity that correlations between any blocks of symbols in
the chain go to zero when the distance between them goes to
infinity. The other consequence of ergodicity is the possibility to
use one random sequence as an equitable representative of the
ensemble of chains and to do averaging over the sequence,
Eq.~(\ref{epsilon-av}), instead of the ensemble averaging.

Below we consider an important class of binary random sequences with
symbols~$a_i$ taking on two values, say $0$~and~$1$, $a_i \in \{0,
1\}$. The conditional probability to find $i$-th element $a_i=1$ in
the \emph{binary} $N$-step Markov sequence depending on $N$
preceding elements $a_{i-N}^{i-1}$ is a set of $2^N$ numbers:
\begin{eqnarray} \label{prob}
&& P(1|a_{i-N}^{i-1})=P(a_{i}=1|a_{i-N}^{i-1}),\nonumber\\[6pt]
&&P(0|a_{i-N}^{i-1}) =1- P(1|a_{i-N}^{i-1}).
\end{eqnarray}
Conditional probability (\ref{prob}) of the binary sequence of
random variables $a_i\in \{0, 1\}$ can be represented exactly as a
\emph{finite} polynomial series:
\begin{eqnarray} \label{prob_series}
&&P(1|a_{i-N}^{i-1}) = \bar{a} + \sum_{r_1=1}^N F_1(r_1)(a_{i-r_1}
- \bar{a}) \nonumber\\
    &&+\sum_{r_1,r_2=1}^N F_2(r_1,r_2)(a_{i-r_1} a_{i-r_2} -
    \overline{a_{i-r_1} a_{i-r_2}}) +
    \ldots  \nonumber\\
    &&+\sum_{r_1,\ldots,r_N=1}^N F_N(r_1,\ldots,r_N)(a_{i-r_1} \ldots
    a_{i-r_N}\nonumber\\
    &&- \,\,\overline{a_{i-r_1} \ldots
    a_{i-r_N}}),
\end{eqnarray}
where the statistical averages $\overline{a_{r_1} \ldots
    a_{r_N}}$ are taken over
sequence~(\ref{epsilon-av}), $F_s$ is the family of \textit{memory
functions} and $\bar{a}$ is the relative average number of unities
in the sequence. The representation of Eq.~(\ref{prob}) in the form
Eq.~(\ref{prob_series}) results from  the simple identical
equalities, $a^2=a$ and $f(a) = a f(1) + (1-a) f(0)$,  for an
arbitrary function $f(a)$ defined on the set $a\in \{0, 1\}$. The
first term in Eq.~(\ref{prob_series}) is responsible for generation
of uncorrelated white-noise sequences. Taking into account the
second term proportional to $F_1(r)$ we can reproduce correctly
correlation properties of the chain up to the second order. In this
case all the correlators of higher orders can be expressed through
the products of the binary correlators. In what follows we will only
use the first two terms, which determine the so-called
\emph{additive} Markov chains~\cite{muya,RewUAMM}. They are in some
sense analogous to autoregressive
models~\cite{Raftery,Berchtold,Chakravarthy}. A particular form of
the conditional probability function of additive Markov chain is the
step-wise memory function,
\begin{eqnarray} \label{prob step}
P(1|k) = \frac{1}{2} + \mu \left(\frac{2k}{N} -1\right).
\end{eqnarray}
The probability $P(1|k)$ of having the symbol $a_i=1$ after $N$-word
$a_{i-N}^{i-1}$ containing $k$ unities, $\,\, k=\sum_{l=1}^{N}
a_{i-l}$, is a linear function of $k$ and is independent of the
arrangement of symbols in the word $ a_{i-N}^{i-1}$. The parameter
$\mu$ characterizes the strength of correlations in the system.

There is a rather simple relation between the memory  function
$F(r)$ (hereafter we will omit the subscript $1$ of $F_1(r)$) and
the pair correlation function of the binary additive Markov chain.
There were suggested two methods for finding the $F(r)$ of a
sequence with a known pair correlation function. The
first~\cite{muya} is based on the minimization of a ``distance''\,
between the Markov chain generated by means of the sought-for memory
function and the initial given sequence of symbols with a known
correlation function. The minimization equation yields the
relationship between the correlation and the memory functions,
\begin{equation} \label{main}
K(r)=\sum\limits_{r'=1}^{N}F(r')K(r-r'), \ \ \ \ r\geq 1.
\end{equation}
where the normalized correlation function $K(r)$ is given by
\begin{equation}\label{def_cor1}
K(r)=\frac{C(r)}{C(0)}, \quad
C(r)=\overline{(a_i-\bar{a})(a_{i+r}-\bar{a})}.
\end{equation}
The second method for deriving Eq.~(\ref{main}) is the completely
probabilistic straightforward calculation~\cite{MUYaG}.

Equation~(\ref{main}), despite its simplicity,  can be analytically
solved only in some particular cases: for one- or two-step chains,
Markov chain with step-wise memory function and so on. To avoid the
difficulties in solving Eq.~(\ref{main}) we suppose that
correlations in the sequence are weak. It means that all components
of the normalized correlation function are small, $|K(r)|\ll 1,
\,|r|\neq 0$, with the exception of $K(0)= 1$. So, taking into
account that in the sum of Eq.~(\ref{main}) the leading term is
$K(0)=1$ and all the others are small, we can obtain an approximate
solution for the memory function in the form of the series
\begin{eqnarray} \label{Series}
F(r)&=&K(r) - \sum_{r'\neq r}^N K(r-r') K(r') \\
&+&\sum_{r'\neq r}^N\sum_{r''\neq r'}^N K(r-r')
K(r'-r'')K(r'')+...\nonumber
\end{eqnarray}
The equation for the conditional probability function in the first
approximation with respect to small functions $|K(r)|\ll 1,
\,|r|\neq 0$, takes the form
\begin{eqnarray} \label{Approx CP}
P(1|a_{i-N}^{i-1})&\simeq &\bar{a} + \sum_{r=1}^N F(r)(a_{i-r} -
\bar{a}) \nonumber\\
&\simeq &\bar{a} + \sum_{r=1}^N K(r)(a_{i-r} - \bar{a}).
\end{eqnarray}

This formula provides a very important tool for constructing a
sequence with a given pair correlation function. Note that the
$i$-independence of function $P(1|a_{i-N}^{i-1})$ guarantees
homogeneity and stationarity of the sequence under consideration;
the finiteness of $N$ together with Eq.~(\ref{ergo_m}) provides its
ergodicity.

The correlation and memory functions are mutually complementary
characteristics of a random sequence in the following sense. The
numerical analysis of a given random sequence enables one to
determine directly the correlation function rather than the memory
function. On the other hand, it is possible to construct a random
sequence using the memory function, but not the correlation one, in
the general case. Therefore, the memory function permits one to get
a deeper insight into the intrinsic properties of the correlated
systems. Equation~(\ref{Approx CP}) shows that in the limit of weak
correlations both functions play the same role.

The concept of additive Markov chain was extensively used earlier
for studying the random sequences with long-range correlations. The
examples and references can be found in~\cite{RewUAMM}.

\section{Differential entropy}

In order to estimate the entropy of infinite stationary sequence
$\mathbb{A}$ of symbols $a_{i}$ one could use the block entropy,
\begin{eqnarray} \label{entro_block}
H_{L}=-\sum_{a_{1},...,a_{L}} P(a_{1}^{L})\log_{2}
P(a_{1}^{L}).
\end{eqnarray}
Here $P(a_{1}^{L}) =P(a_{1},\ldots,a_{L})$ is the probability to
find the $L$-word $a_{1}^{L}$ in the sequence. The differential
entropy, or entropy per symbol, is given by
\begin{eqnarray} \label{entro_diff}
h_{L}=H_{L+1} - H_{L},
\end{eqnarray}
and specifies the degree of uncertainty of the $(L+1)$-th symbols
observing if the preceding $L$ symbols are specified. The source
entropy (or Shannon entropy) is the differential entropy at the
asymptotic limit, $h=\lim_{L \rightarrow \infty}h_{L}$. This
quantity  measures the average information per symbol if {\it all}
correlations, in the statistical sense, are taken into account.

The differential entropy $h_L$ can be presented in terms of the
conditional probability function. To show this we have to rewrite
Eq.~\eqref{entro_block} for the block of length $L+1$,  expressing
$P(a_{1}^{L+1})$ via the conditional probability, and after a bit of
algebra we obtain %%
%%%
\begin{eqnarray} \label{Entro_Bin}
h_L=\!\!\!\sum_{a_{1},...,a_{L}=0,1} \!\!\!P(a_{1}^{L})
h(a_{L+1}|a_{1}^{L}) = \overline{ h(a_{L+1}|a_{1}^{L})}.
\end{eqnarray}
Here $h(a_{L+1}|a_{1}^{L})$ is the conditional (not averaged)
entropy or the amount of information contained in the $(L+1)$-th
symbol of the sequence conditioned on $L$ previous symbols,
\begin{eqnarray}
   h(a_{L+1}|a_{1}^{L}) = - \!\!\!\sum_{a_{L+1}=0,1}\!\!\!
P(a_{L+1}|a_{1}^{L})\log_2 P(a_{L+1}|a_{1}^{L}).
    \label{siL}
\end{eqnarray}
So, the differential entropy $h_L$ of a random sequence is presented
as a generalization of the standard conditional entropy $H=-\sum_{A}
P(A) \sum_{B} P(B|A)\log_2 P(B|A)$ to the multi-symbol event $
a_{1}^{L}$.

The conditional probability $P(1|a_{i-L}^{i-1})$ at $L < N$,
\begin{eqnarray} \label{p_i(L)}
P(1|a_{i-L}^{i-1}) \simeq \bar{a} + \delta; \quad \delta =
\sum_{r=1}^L F(r)(a_{i-r} -\bar{a}),
\end{eqnarray}
can be obtained in the first approximation in parameter $\delta$
from Eq.~\eqref{Approx CP} by means of a simple probabilistic
reasoning.

Taking into account the weakness of correlations, $|\delta| \ll \min
[\overline{a}, (1-\overline{a})]$, one can expand the right-hand
side of Eq.~\eqref{siL} in Taylor series up to the second order in
$\delta$, $h(a_{L+1}|a_{1}^{L}) = h_0 + (\partial
h/\partial\overline{a})_{|\delta=0}\delta + (1/2)(\partial^2
h/\partial\overline{a}^2)_{|\delta=0}\delta^2$, where the
derivatives are taken at the ``equilibrium point''
$P(1|a_{i-L}^{i-1}) = \bar{a}$ and $h_0$ is the entropy of
uncorrelated sequence,
\begin{eqnarray}
h_0=-\bar{a}\log_{2}(\bar{a}) - (1-\bar{a})\log_{2}(1-\bar{a}).
\end{eqnarray}
Upon using Eq.~\eqref{Entro_Bin} for averaging
$h(a_{L+1}|a_{1}^{L})$ and in view of $ \overline{\delta} =0$, the
differential entropy of the sequence becomes
\begin{equation}\label{Entro_Markov2}
h_L= \left\{\begin{array}{l} h_{L \leq N}= h_0 -
\dfrac{1}{2\ln2}\sum_{r=1}^L F^2(r), \\
 h_{L>N}=h_{L=N}.
\end{array}
\right.
\end{equation}
If the block length exceeds the memory length, $L>N$, the
conditional probability $P(1|a_{i-L}^{i-1})$ depends only on $N$
previous symbols, see Eq.~(\ref{def_mark}). Then, it is easy to show
from~\eqref{Entro_Bin} that the differential entropy remains
constant at $ L \ge N$. The second line of Eq.~\eqref{Entro_Markov2}
is consistent  with the first one because in the first approximation
in $\delta$ the correlation function vanishes at $L>N$ together with
the memory function.  The final expression, the main result of the
paper,  for the differential entropy of the stationary ergodic
binary weakly correlated random sequence is
\begin{equation}\label{EntroMain}
h_L=  h_0 - \frac{1}{2\ln2}\sum_{r=1}^L K^2(r).
\end{equation}

It follows from Eq.~(\ref{EntroMain}) that the additional correction
to the entropy $h_0$ of uncorrelated sequence is the negative and
monotonously decreasing function of $L$. It is the anticipated
result -- the correlations reduce entropy. The conclusion is not
sensitive to the sign of correlations: persistent correlations,
$K>0$, describing the ``attraction'' of symbols of the same kind,
and anti-persistent correlations, $K<0$, corresponding to the
attraction between ``0'' and ``1'', provide the corrections of the
same negative sign.

If the correlation function is constant at $1 \leq r \leq N$, the
entropy is a linear decreasing function of the argument $L$ up to
the point $N$; the result is coincident with that obtained
in~\cite{DMBUYa} (in the limit of weak correlations) for the Markov
chain model with step-wise memory function~(\ref{prob step}).
\begin{figure}[h!]
\center\includegraphics[width=0.48\textwidth]{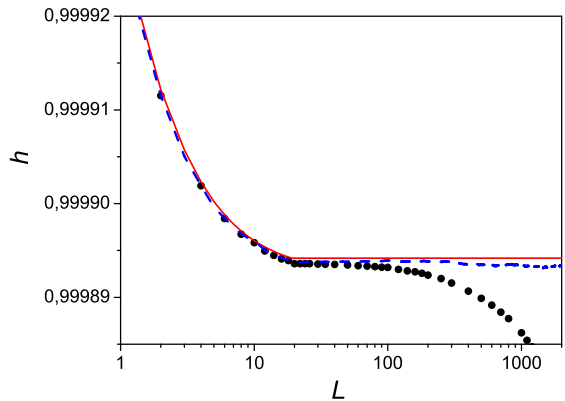}
\caption{The differential entropy vs the length $L$. The solid line
is the analytical result, Eq.~(\ref{EntroMain}), for correlation
function $K(r)=0.01/r^{1.1} $, whereas the dots correspond to direct
evaluation of the same Eq.~(\ref{EntroMain}) for the numerically
constructed sequence (of the length $M=10^8$ and the cut-off
parameter $R_c=20$) by means of conditional probability
function~(\ref{Approx CP}) and the numerically evaluated correlation
function $K(r)$ of the constructed sequence. The dashed line is the
differential entropy with fluctuation correction described by
Eq.~(\ref{EntroFin}). } \label{N_c20}
\end{figure}

As an illustration of result (\ref{EntroMain}), in Fig.~\ref{N_c20}
we present the plot of the differential entropy versus the length
$L$. Both numerical and analytical results (the dotted and solid
curves) are presented for the power-law correlation function
$K(r)=0.01/r^{1.1} $. The cut--off parameter $R_c$ of the power-law
function for numerical generation of the sequence, coinciding with
the memory length of the chain, is $20$. The good agreement between
the curves at $L<R_c$ is the manifestation of adequateness of the
additive Markov chain approach for studying the entropy properties
of random chains.

\section{Finite random sequences}

The relative average number of unities $\bar{a}$, correlation
functions and other statistical characteristics of random sequences
are deterministic quantities in the limit of their infinite lengths
only. It is a direct consequence of the law of large numbers. If the
sequence length $M$ is finite, the set of numbers $a_1^M$ cannot be
considered anymore as ergodic sequence. In order to restore its
status we have to introduce the \emph{ensemble} of finite sequences
$\{a_1^M\}_p, p \in \mathbb{N} =0,1,2,...$. Yet, we would like to
retain the right to examine \emph{finite} sequences %even if
%approximately
by using a single finite chain. So, for a finite chain
we have to replace definition~(\ref{def_cor1}) of the correlation
function by the following one,
%
%\begin{equation}\label{CorFin}
%C_M(r)=\frac{1}{M-r}
%\sum_{i=0}^{M-r-1}(a_i-\bar{a})(a_{i+r}-\bar{a}),
%\bar{a}=\frac{1}{M}\sum_{i=0}^{M-1}a_i.
%\end{equation}
%
\begin{eqnarray} \label{CorFin}
C_M(r)&=&\frac{1}{M-r}
\sum_{i=0}^{M-r-1}(a_i-\bar{a})(a_{i+r}-\bar{a}),\nonumber \\
\bar{a}&=&\frac{1}{M}\sum_{i=0}^{M-1}a_i.
\end{eqnarray}

Now the correlation functions and $\bar{a}$ are random quantities
which depend on the particular realization of the sequence $a_1^M$.
Their fluctuations can contribute to the entropy of finite random
chains even if the correlations in the random sequence are absent.
It is well known that the order of relative fluctuations of additive
random quantity (as, e.g., the correlation function
Eq.~(\ref{CorFin})) is $1/\sqrt{M}$.

Below we give more rigorous justification of this explanation and
show its applicability to our case. Let us present the correlation
function $C_M(r)$ as the sum of two components,
\begin{equation}\label{CorrelSquar}
C_M(r)= C(r)+C_{f}(r),
\end{equation}
where the first summand $C(r)=\lim_{M\rightarrow\infty} C_M(r)$ is
the correlation function determined by Eqs.~(\ref {def_cor1}) and
(\ref {CorFin}), obtained by averaging over the sequence with
respect to index $i$, enumerating the elements $a_{i}$ of sequence
$\mathbb{A}$; and the second one, $C_{f}(r)$, is a
fluctuation--dependent contribution. Function $C(r)$ can be also
presented as the ensemble average $C(r)=\langle C_M(r) \rangle$ due
to the ergodicity of the sequence.

Now we can find a relationship between variances of $C_M(r)$ and
$C_{f}(r)$. Taking into account Eq.~(\ref {CorrelSquar}) and the
properties $\langle C_f(r) \rangle =0$ at $r\neq 0$ and
$C(r)=\langle C_M(r) \rangle$  we have
\begin{equation}\label{CorrelSquar1}
\langle C^2_{M}(r) \rangle = C^2(r) + \langle C^2_{f}(r) \rangle.
\end{equation}

The mean fluctuation of squared correlation function $C^2_f(r)$ is
%
%\begin{equation}\label{Fluct}
%\langle C^2_f(r)\rangle =
%\frac{1}{(M-r)^2}\langle\sum_{n,m=0}^{M-r-1}(a_n-\bar{a})(a_{n+r}-\bar{a})
%(a_m-\bar{a})(a_{m+r}-\bar{a})\rangle.
%\end{equation}
\begin{eqnarray}\label{Fluct}
&&\langle C^2_f(r)\rangle = \\[6pt]
&&\!\!\!
\frac{1}{(M-r)^2}\langle\!\!\!\sum_{n,m=0}^{M-r-1}\!\!\!(a_n-\bar{a})(a_{n+r}-\bar{a})
(a_m-\bar{a})(a_{m+r}-\bar{a})\rangle. \nonumber
\end{eqnarray}

Taking into account that only the terms with $n=m$ give nonzero
contribution to the result and neglecting correlations between
elements $a_n$,
\begin{eqnarray}\label{Fluct2}
& \langle & \sum_{n,m=0}^{M-r-1}(a_n-\bar{a})(a_{n+r}-\bar{a})
(a_m-\bar{a})(a_{m+r}-\bar{a})\rangle\\ \nonumber &=&
\sum_{n=0}^{M-r-1}\langle(a_n-\bar{a})^2\rangle\langle(a_{n+r}-\bar{a})^2
\rangle=(M-r)\,\,C^2_f(0) .
\end{eqnarray}

we obtain for the normalized correlation function

\begin{equation}\label{FluctFin}
\langle K^2_f(r) \rangle = \frac{\langle C^2_f(r)\rangle}{C^2_f(0)},
\,\, \langle K^2_f(r) \rangle = \frac{1}{M-r}\simeq \frac{1}{M}.
\end{equation}

Note that Eq.~(\ref{FluctFin}) is obtained by means of averaging
over the ensemble of chains. This is the shortest way to obtain the
desired result. At the same time, for numerical simulations we have
used only the averaging over the chain as is seen from
Eq.~(\ref{CorFin}), where the summation over sites $i$ of the chain
plays the role of averaging.

Note also that the different symbols $a_i$ in Eq.~(\ref{Fluct}) are
correlated. It is possible to show that contribution of their
correlations to $\langle K^2_f(r) \rangle$ is of order $R_{c}/M^2\ll
1/M$.

The fluctuating part of entropy, proportional to $\sum_{r=1}^L
K^2_f(r)$, should be subtracted from Eq.~(\ref{EntroMain}), which is
only valid  for the infinite chain.
Thus, Eqs.~(\ref{CorrelSquar1}) and~(\ref{FluctFin}) yield the
differential entropy of the \emph{finite} binary weakly correlated
random sequences
\begin{equation}\label{EntroFin}
h_L=  h_0 - \frac{1}{2\ln2}\left[\sum_{r=1}^L K_M ^2(r)-\ln
\frac{M}{M-L}\right].
\end{equation}

It is clear that in the limit $M\rightarrow\infty$ this function
transforms into Eq.~(\ref{EntroMain}).  When $L\ll M$, the last term
in rhs of Eq.~(\ref{EntroFin}) takes the form $L/M$ and describes
the linearly decreasing entropy.
%in the inset of Fig.~\ref{Weak_cor}.

The squared correlation function $K^2_M(r)$ is normally a decreasing
function of $r$, whereas function $K^2_f(r)$ is an increasing one.
Hence, the terms $\sum_{r=1}^L K_M ^2(r)$ and $\ln [M/(M-L]$ being
concave and convex functions, respectively, describe the competitive
contributions to the entropy. It is not possible to analyze all
particular cases of their relationship. Therefore we indicate here
the most interesting ones taking in mind monotonically decreasing
correlation functions. An example of such type of function,
$K(r)=a/r^{b}, \, a>0, \,b \geq 1 $, was considered above.

If the correlations are extremely small and compared with the
inverse length $M$ of the sequence, $K_M ^2(1) \sim 1/M$, the
fluctuating part of the entropy exceeds the correlation part nearly
for all values of $L>1$.

With the increasing of $M$ (or correlations), when the inequality
$K_M ^2(1)> 1/M$ is fulfilled, there is at list one point where the
contribution of fluctuation and correlation parts of the entropy are
equal. For monotonically decreasing function $K(r)$ there is only
one such point. Comparing the functions in square brackets in
Eqs.~(\ref{EntroFin}) we find that they are equal at some $L =
R_{s}$, which hereafter will be referred to as a stationarity
length. If $L \ll R_s$, the fluctuations of the correlation function
are negligibly small with respect to its magnitude, hence the finite
sequence may be considered as quasi-stationary one. At $ L \sim
R_{s}$ the fluctuations are of the same order as the genuine
correlation function $K^2(r)$. Here we have to take into account the
fluctuation correction due to the finiteness of the random chain. At
$ L > R_{s}$ the fluctuating contribution exceeds the correlation
one.

\begin{figure}[h!]
{\center\includegraphics[width=0.48\textwidth]{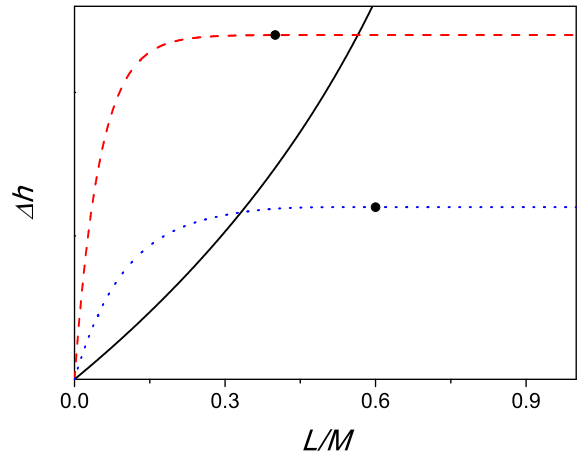}}
\caption{The dotted and dashed lines are $M$-independent
contributions to the entropy, $ \sum_{r=1}^L K_M ^2(r)\,/ \, 2\ln
2$, see Eq.~(\ref{EntroFin}), for two different memory lengths
marked by two solid dots. Both lines correspond to exponential
correlator $K(r) \propto \exp(-r/r_0)$. For the dashed line $r_0 =
0.1 M$ and correlation length is $r_c = 0.4 M$. The dotted line
represents a sequence with $r_0 = 0.2 M$ and $r_c = 0.6 M$. The
solid line is the fluctuation correction $ \ln [M/(M-L)]\,/\, 2\ln
2$. } \label{two_corr}
\end{figure}

The other important parameter of the random sequence is the memory
length $N$. If the length $N$ is less than $R_{s}$, we have no
difficulties to calculate the entropy of the finite sequence, which
can be considered as quasi-stationary. This case is illustrated in
Fig.~\ref{N_c20} where the good agreement between the analytical and
numerical curves at $L<R_c$ is clearly seen. If the memory length
exceeds the stationarity length, $R_{s} \lesssim N$, we have to take
into account the fluctuation correction to the entropy. The entropy
with this correction is shown in Fig.~\ref{N_c20} by the dashed
line. Two types of different relationships between memory length $N$
and stationarity length $R_{s}$ are shown in Fig.~\ref{two_corr}.
Note that at $L>N$ the entropy does not change. Two solid points in
the figure correspond to the equality $L=N$.

\section{Entropy of DNA sequences}

It is known that any DNA text is written by four ``characters'',
specifically by adenine (A), cytosine (C), guanine (G), and thymine
(T). Therefore, there are three nonequivalent types of the DNA text
mapping onto one-dimensional binary sequences of zeros and unities.
The first of them is the so-called purine-pyrimidine rule, \{A,G\}
$\rightarrow$ 0, \{C,T\} $\rightarrow$ 1. The second one is the
hydrogen-bond rule, \{A,T\} $\rightarrow$ 0, \{C,G\} $\rightarrow$
1. And, finally, the third is \{A,C\} $\rightarrow$ 0, \{G,T\}
$\rightarrow$ 1.

\begin{figure}[h!]
{\center\includegraphics[width=0.42\textwidth]{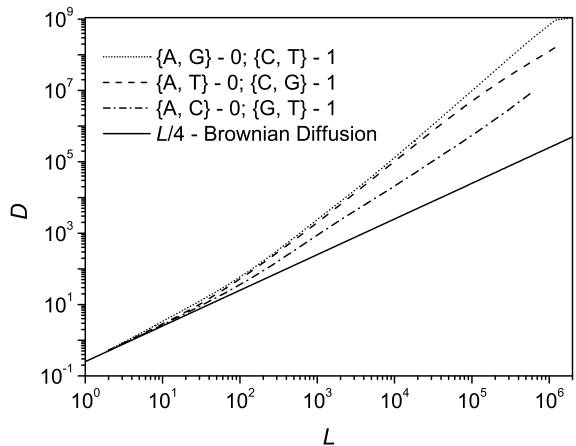}}
\caption{The dependence $D(L)$ for the coarse-grained DNA text of
\textit{Bacillus subtilis, complete genome} ~\cite{ftp}, for three
nonequivalent kinds of mapping. Dotted, dashed, and dash-dotted
lines correspond to the purine-pyrimidine mapping, \{A,G\}
$\rightarrow$ 0, \{C,T\} $\rightarrow$ 1; hydrogen-bond rule
mapping, \{A,T\} $\rightarrow$ 0, \{C,G\} $\rightarrow$ 1; and
\{A,C\} $\rightarrow$ 0, \{G,T\} $\rightarrow$ 1, respectively. The
solid line describes the non-correlated Brownian diffusion,
$D(L)=L/4$.} \label{f12}
\end{figure}

In order to understand which kind of mapping is more appropriate for
calculating the entropy, we consider all three kinds of
mapping~\cite{UYa}. As an example, the variance
$D(L)=\overline{k^{2}}-\overline{k}^{2},
k_{i}(L)=\sum\limits_{l=1}^{L}a_{i+l}$ for the coarse-grained text
of \textit{Bacillus subtilis, complete genome}~\cite{ftp}, is
displayed in Fig.~\ref{f12} for all possible types of mapping. The
different kinds of mapping reveal and emphasize various types of
physical attractive correlations between the nucleotides in the DNA,
such as the strong purine-purine and pyrimidine-pyrimidine
persistent correlations (the upper curve), and the correlations
caused by the weaker attraction A$\leftrightarrow$T and
C$\leftrightarrow$G (the middle curve). In what follows we will use
the purine-pyrimidine coarse-grained mapping, which corresponds to
the strongest correlations.
\begin{figure}[h!]
\center\includegraphics[width=0.48\textwidth]{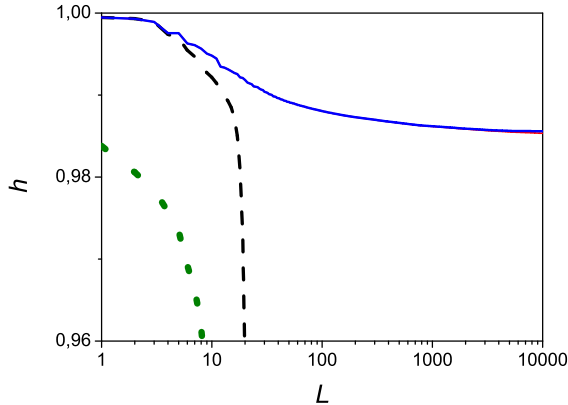}
\caption{Differential entropy $h$ vs length $L$ for R3 chromosome of
\emph{Drosophila melanogaster} DNA of length $M\simeq2.7\times
10^7$. The solid line is obtained by using Eq.~(\ref{EntroMain})
with numerically evaluated correlation function
Eq.~(\ref{def_cor1}). The dashed line is the differential entropy,
Eqs.~(\ref{entro_block}) and (\ref{entro_diff}), plotted by using
the numerical estimation of probability $P(a_{1},\ldots,a_{L})$ of
the $L$-blocks occurring in the same  sequence. The dots are the
differential entropy (normalized by division by 2) of the same
sequence, Eqs.~(\ref{entro_block}) and (\ref{entro_diff}), without
coarse-graining, i.e., for four-letter DNA sequence. }
\label{Graph2}
\end{figure}

In order to evaluate the entropy of DNA sequence using
Eq.~(\ref{EntroMain}) at first we have to calculate the normalized
correlation function given by Eq.~(\ref{def_cor1}), where each
random variable $a_i$ after mapping takes on the values $0$ or $1$.
The result of such calculation for R3 chromosome of \emph{Drosophila
melanogaster} DNA of length $M\simeq2.7\times 10^7$ is shown in
Fig.~\ref{Graph2} by the solid line. The abrupt deviation of the
dashed line from the upper curves at $L\sim 10$ is the result of
violation of inequality~(\ref{Strong2}) and the manifestation of
rapidly growing errors in the entropy estimation by using the
probability $P(a_{1},\ldots,a_{L}) $ of the $L$-blocks occurring.
The dotted curve shows that the violation of strong
inequality~(\ref{Strong2}) for four-letter sequence begins at
smaller value of $L$ than for two-letter (binary) sequence.
\begin{figure}[h!]
\center\includegraphics[width=0.48\textwidth]{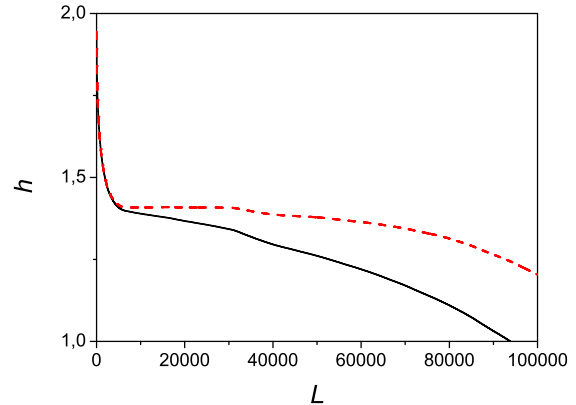}
\caption{The differential entropy of \emph{Homo sapiens} chromosome
Y, locus NW 001842422 vs length $L$. The solid line is obtained by
using the equation similar to Eq.~(\ref{EntroMain}) with numerically
evaluated correlation functions. The dashed line is the entropy with
the fluctuation correction. } \label{h_NW_001842422}
\end{figure}

The theory of additive Markov chains presented here can be applied
to the chains with $d$-valued space of states. In our case $d=4$.
Using the formula similar to Eq.~(\ref{EntroFin}) we evaluate the
entropy for \emph{Homo sapiens} chromosome Y, locus NW 001842422.
The result of calculation is shown in Fig.~\ref{h_NW_001842422}. It
is clearly seen that the entropy in interval $7\times 10^3<L<3\times
10^4$ takes on the constant value, $h_{L}\simeq 1.41$. It means that
for $L>7\times 10^3$ {\it all} correlations, in the statistical
sense, are taken into account, or, in other words, the memory length
of the \emph{Homo sapiens} chromosome Y is of the order of $10^4$.
At $L>3\times 10^4$ the entropy evidently should be constant as
well. The presented deviation is the consequence of many different
reasons such as the nonadditivity of the sequence under study, the
violation of supposed weakness of correlations, and many others.

We believe that, along with the memory length, the asymptotic value
of the entropy $h_{L}$ at $L\rightarrow \infty$ (the Shannon source
entropy) can be the important characteristics of the living species.

\section{Conclusion and perspectives}

1. This paper is the first application of the theory based on the
additive Markov chains approach for describing the DNA sequences. It
is evident that we need a more systematic and extensive study of the
real biological sequences.

2. We have supposed that correlations are weak. However, our
preliminary study evidences that when correlations are not weak, a
strong short-range part in the interaction of symbols changes in
Eq.~(\ref{EntroMain}) the numerical coefficient before the term
$\sum_{r=1}^L K^2(r)$ at $L\rightarrow\infty$.

3. Our consideration can be generalized to the Markov chain with the
infinite memory length $N$. In this case we have to impose a
condition on the decreasing rate of the correlation function and the
conditional probability function at $N\rightarrow \infty$. Another
generalization, which may be important for biological applications
~\cite{Raftery,Cocho,Tavare,Borodovsky}, is the non-homogenous
Markov chains. In this case the conditional probability function $P$
has to be the function of the position $i$ of symbol $a_i$,
\begin{eqnarray}\label{non_hom_mark}
P=P(a_i=a|i, a_{i-N},\ldots,a_{i-2},a_{i-1}).
\end{eqnarray}

4. It would be interesting to compare the result obtained in our
work with that of the Lempel-Zive approach~\cite{Cover} and the
hidden Markov chain model~\cite{Seifert}.

5. In this paper we have considered the random sequences with the
binary space of states, but almost all results can be generalized to
non-binary sequences.

\acknowledgments We are grateful for the very helpful and fruitful
discussions with A.~A.~Krokhin, S.~V.~Denisov, S.~S.~Apostolov,
Z.~A.~Mayzelis, G.~M.~Pritula, and Yu.~V.~Tarasov.

%%%%%%%%%%%%%%%%%%%%%%%%%%%%%%%%%%%%%%%%%%%%%%%%%%%%%%%%%%%%%%%%%%


\begin{thebibliography}{20}

\bibitem{Buldyrev} Buldyrev,~S.V. et al, 1995. Long-range correlation properties of coding and noncoding DNA sequences: GenBank analysis. Phys.~Rev.~E. 51, 5084.

\bibitem{Almir} Almirantis,~Y., Provata,~A., 1999. Long- and Sort-Range Correlations in Genome Organisation. J.~Stat.~Phys., 97, 233.

\bibitem{mad} Madigan,~M.T., Martinko,~J.M., Parker,~J., 2002. Brock Biology of Microorganisms, Prentice Hall.

\bibitem{Eren} Ehrenfest,~P., Ehrenfest,~T, 1911. Encyklop\"{a}die der Mathematischen Wissenschaften, Berlin:~Springer.

\bibitem{Lind} Lind,~D., Marcus,~B., 1995. An Introduction to Symbolic Dynamics and Coding,
Cambridge University Press.

\bibitem{Shan} Shannon,~C.E., Weaver,~W., 1949. The Mathematical Theory of Communication, University of Illinois Press.

\bibitem{Cover} Cover,~T.M., Thomas,~J.A., 1991. Elements of Information Theory, Wiley, New York.

\bibitem{muya} Melnyk,~S.S., Usatenko~O.V., Yampol'skii,~V.A., 2006. Memory functions of the additive Markov chains: applications to complex dynamic systems. Physica~A, 361, 405.

\bibitem{RewUAMM} Usatenko~O.V., Apostolov,~S.S., Mayzelis~Z.A., Melnik,~S.S., 2010. Random finite-valued dynamical systems: additive Markov chain approach, Cambridge Scientific Publisher, Cambridge.

\bibitem{Raftery} Raftery,~A., 1985. A model for high-order Markov chains. Journal of Royal Statistical Society B, 47, 528-539.

\bibitem{Ching} Ching,~W.K., Fung,~E.S., Ng,~M.K., 2004. Higher-order Markov chain models for categorical data sequence. Naval Research Logistics, 51, 557-574.

\bibitem{Li} Li,~W.K., Kwok,~M.C.O., 1990. Some results on high order Markov chain models. Communications in Statistics - Simulation and Computation, 19, 363-380.

\bibitem{Cocho} Cocho,~J.A. et al, Bacterial genomes lacking long-range correlations may not be modeled by low-order Markov chains, this issue.

\bibitem{Seifert} Seifert,~M., Gohr,~A., Strickert,~M., Grosse,~I., 2012. Parsimonious higher-order hidden Markov models for improved array-CGH analysis with applications to Arabidopsis thaliana, PLoS Computational Biology, 8, e1002286.

\bibitem{shir} Shiryaev,~A.N., 1996. Probability, Springer, New York.

\bibitem{Berchtold} Berchtold,~A., 1995. Autoregressive modelling of Markov chains, in: Statistical Modelling, Lecture Notes in Statistics, vol 104, Springer, pp.19-26.

\bibitem{Chakravarthy} Chakravarthy,~N., Spanias,~A., Iasemidis,~L.D., Tsakalis,~K., 2004. Autoregressive modeling and feature analysis of DNA sequences, EURASIP J Applied Signal Processing, 1, 13-28.

\bibitem{MUYaG} Melnyk,~S.S., Usatenko,~O.V., Yampol'skii,~V.A., Golick,~V.A., 2005. Competition between two kinds of correlations in literary texts, Phys.~Rev.~E, 72, 026140.

\bibitem{DMBUYa} Denisov,~S.V., Melnik,~S.S., Borisenko,~A.A., Usatenko,~O.V., Yampolsky,~V.A., Entropy of complex symbolic sequences: Exact results; to be published.

\bibitem{UYa} Usatenko,~O.V., Yampol'skii,~V.A., 2003. Binary N -Step Markov Chains and Long-Range Correlated Systems, Phys.~Rev.~Lett., 90, 110601.

\bibitem{Tavare} Raftery,~A., Tavare,~S., 1994. Estimation and modelling repeated patterns in high order Markov chains with the mixture transition distribution model, Applied Statistics, 43, 179-199.

\bibitem{Borodovsky} Borodovsky,~M., Peresetsky,~A., 1994. Deriving Non-homogeneous Markov Chain Models by Cluster Analysis Algorithm Minimizing Multiple Alignment Entropy, Computers and Chemistry, 18, 259-268.

\bibitem{ftp} ftp:$//$ftp.ncbi.nih.gov$/$genomes$/$bacteria$/$bacillus\_subtilis  $/$NC\_000964.gbk.
\end{thebibliography}
\end{document}